\documentclass[prl,twocolumn,noshowpacs,preprintnumbers,amsmath,amssymb]{revtex4}

\usepackage{graphicx}
\usepackage{dcolumn}
\usepackage{bm}
\usepackage{color}
\usepackage{soul}
\usepackage{multirow}

\begin{document}

 \preprint{APS/123-QED}
\title{Achieving Identical Stored Energy in Cascaded Collisional Quantum Battery Charging: Analytical Results}


\author{Longyan Gong  $^{1,2}$}
\thanks{Email address: lygong@njupt.edu.cn.}
\author{Jing Zhang $^{1,2}$}
\author{Yongtao Li $^{1,2}$}
\affiliation{
 $^{1}$College of Science, Nanjing University of Posts and Telecommunications, Nanjing, 210003, China\\
 $^{2}$Jiangsu Provincial Engineering Research Center of Low Dimensional Physics and New Energy, Nanjing, 210003, China\\
 }
\date{\today}
\begin{abstract}
Charging uniformity across distinct trials is an essential requirement for reliable quantum batteries. Nevertheless, intrinsic quantum stochasticity poses a fundamental challenge to this goal. This work investigates a two-qubit quantum battery within the framework of cascaded collision models. We propose an adaptive measurement on the charger after each collision, yielding two conditional battery states of equal expected stored energy. For chargers prepared in the excited state, we analytically prove that complete suppression of trajectory-dependent stored-energy fluctuations is attainable; for chargers prepared in general superposition states, our theory reveals that full suppression is fundamentally prohibited owing to battery-phase-dependent recursive dynamics. Numerical simulations are in excellent quantitative agreement with theoretical predictions. These findings provide a feasible pathway toward high-uniformity quantum-battery charging.
\end{abstract}

\maketitle
\emph{Introduction}---Quantum batteries, first proposed by Alicki and Fannes in 2013~\cite{AL13}, exploit quantum effects to transfer, store and release energy. On the one hand, they may serve as energy-supply units compatible with quantum architectures such as quantum computers, quantum sensors and quantum simulators~\cite{AU22,CA24}. On the other hand, they provide a framework for quantum thermodynamics to uncover new principles for energy manipulation and examine the associated quantum advantages~\cite{FE26}.
A series of proof-of-principle demonstrations have validated the core operational characteristics and practical feasibility of quantum batteries~\cite{CA25}. For instance, experiments have been performed on NMR star-topology spin systems~\cite{JO22}, superconducting qubit processors~\cite{YU24}, photonic quantum switches~\cite{ZG23}, and organic optical microcavities~\cite{QU22,HY26}. Over the past decade, quantum batteries have attracted growing interest and emerged as promising candidates for advanced energy storage and quantum-state manipulation~\cite{CA24,FE26,CA25,KH26,TA26}.

To investigate the working mechanisms of quantum batteries, many models have been extensively studied~\cite{CA24,FE26,KH26}, for instance, the Dicke~\cite{FE18}, Tavis-Cummings~\cite{YA24}, Rabi~\cite{FE18}, spin-chain~\cite{LE18}, Sachdev-Ye-Kitaev~\cite{RO20}, and Su-Schrieffer-Heeger models~\cite{ZH26}. Using these models, researchers assess the performance including stored energy, charging power and ergotropy~\cite{YA24,FE18,LE18,RO20,ZH26}, evaluate the detrimental impacts of environmental noise and decoherence~\cite{FE26,ZH26}, and clarify the physical origins of quantum advantage~\cite{FE18,RO20}.

Generally, the total Hamiltonian can be expressed as~\cite{FE26}
\begin{equation}
H(t)=H_\text{a}+H_\text{b}+H_\text{a-b}(t).\label{EQ1}
\end{equation}
Here, the subscripts a and b refer to the charger and battery, respectively, with \(H_\text{a-b}(t)\) describing the time-dependent coupling between them. During the charging process, the coupling is turned on at $t=0$. This interaction lasts for a finite time interval $T$, referred to as the charging time, and the coupling is switched off at $t=T$. In general, the battery resides in a mixed state after it is disconnected from the charger. This state is described by the reduced density matrix \(\rho_\text{b}(T) = \text{tr}_\text{a}\left[\rho(T)\right]\), which is obtained by tracing out the charger's degrees of freedom from the joint density matrix $\rho(T)$. The stored energy \(E(T)=\text{tr}\left[H_\text{b}\rho_\text{b}(T)\right]-\text{tr}\left[H_\text{b}\rho_\text{b}(0)\right]\)~\cite{FE18,YA24,LE18,RO20,ZH26},
the charging power
\(P(T)=\frac{E(T)}{T}\)~\cite{FE18,LE18,RO20}, and the energy fluctuation \(
\delta(T)=\sqrt{\text{tr}\left(H_\text{b}^2\rho_\text{b}(T)\right)-\left[\text{tr}\left(H_\text{b}\rho_\text{b}(T)\right)\right]^2}\)~\cite{FH18,CR20,DO26}.

Charging uniformity across different trials is a key requirement for reliable devices. This means nearly identical stored energy among distinct realizations. We define stored-energy fluctuations as the trajectory-to-trajectory standard deviation of the final energy stored in the battery just after charging, namely the fluctuations arising from the stochasticity of individual charging realizations rather than the quantum uncertainty of a single state. Recently, Mitchison \textit{et al.} tackled this problem and they carried out stochastic master-equation simulations of quantum trajectories~\cite{MI21}. Each trajectory corresponds to one possible realization of the charging process. They tried to suppress stored-energy fluctuations using instantaneous and delayed linear feedbacks.  Even so, stored-energy fluctuations remain significant. This raises a natural question: can one find a suitable approach to fully suppress such fluctuations?

Alternatively, quantum collision models describe system dynamics through repeated discrete interactions between subsystems; such schemes are also known as repeated interaction schemes~\cite{CI22}. By measuring each ancilla immediately after its collision with the target system, one obtains the conditional post-measurement state of the target system; repeating this process yields individual stochastic evolution trajectories. In the weak-coupling and continuous-time limits, the resulting conditional dynamics agree with those of the stochastic master equation. Benefiting from this unique trajectory-resolved capability, collisional frameworks have been widely adopted in recent years to explore the charging dynamics of quantum batteries, including quantum charging speed-up~\cite{SE21,CH22} and non-Markovian environmental effects~\cite{MO23}. Although stored-energy fluctuations have been mentioned within collision models~\cite{SE21}, to the best of our knowledge, no study has demonstrated the complete suppression of such fluctuations.

In this Letter, we adopt a two-qubit cascaded collisional quantum-battery framework and develop an adaptive-measurement protocol for the charger. With this protocol, complete elimination of stored-energy fluctuations can be realized for excited-state chargers, while substantial fluctuation suppression is achieved for superposition-state chargers. Our results offer a feasible pathway toward stable and uniform quantum-battery charging.

\emph{Model}---We investigate the simplest yet nontrivial charger--battery configuration,
where both the charger and battery are individual two-level qubits
~\cite{DO26,AN18,FA19}. The total system Hamiltonian, given in
Eq.~\eqref{EQ1}, consists of the following components,
\begin{equation}
\begin{aligned}
&H_a=\omega_a\,\sigma^{\dagger}_a \sigma_a,
H_b=\omega_b\,\sigma^{\dagger}_b \sigma_b,\\
&H_{a-b}=g\big(\sigma_a\,\sigma^{\dagger}_b+\sigma^{\dagger}_a\,\sigma_b\big).
\end{aligned}
\label{EQ2}
\end{equation}
Here \(\omega_m\) is the transition frequency of the two-level qubit \(m\) with \(m=a\) or \(b\), and the ladder operators are defined by
\(\sigma_m=|0\rangle_m\langle 1|\) and \(\sigma^{\dagger}_m=|1\rangle_m\langle 0|\),
where \(|0\rangle_m\) and \(|1\rangle_m\) denote the ground and excited states,
respectively. Without loss of generality, the ground state is taken as the zero-energy reference, so that
the excited state \(|1\rangle_m\) has energy \(\omega_m\). The parameter \(g\) denotes the coupling strength between
the charger and the battery. We adopt this model for two reasons: it admits an exact solution under free evolution and yields universal features independent of specific experimental implementations~\cite{AN18}.

\emph{Collision, i.e., unitary evolution with interaction}---At the beginning of a single realization of the charging process, the charger and the battery are prepared in pure states \(|\phi_a(0)\rangle\) and \(|\phi_b(0)\rangle\). The joint state,  \(|\psi(0)\rangle=|\phi_a(0)\rangle\otimes|\phi_b(0)\rangle\), can be generally written as \(|\psi(0)\rangle=c_{00}|00\rangle+c_{01}|01\rangle+c_{10}|10\rangle+c_{11}|11\rangle\). Obviously, it is a separable state, whose time evolution is given by
\(|\psi(t)\rangle=e^{-iHt}|\psi(0)\rangle\). This takes the explicit form
\begin{equation}
|\psi(t)\rangle = c_{00}(t)|00\rangle + c_{01}(t)|01\rangle + c_{10}(t)|10\rangle + c_{11}(t)|11\rangle,
\label{EQ3}
\end{equation}
 where
\begin{equation}
\label{EQ4}
\begin{cases}
c_{00}(t)=c_{00},\\
c_{01}(t)=e^{-i\bar\omega t}\left[
c_{01}\left(\cos\Omega t -i\frac{\Delta}{2\Omega}\sin\Omega t\right)
-\,i c_{10}\frac{g}{\Omega}\sin\Omega t
\right],\\
c_{10}(t)=e^{-i\bar\omega t}\left[
-\,i c_{01}\frac{g}{\Omega}\sin\Omega t
+c_{10}\left(\cos\Omega t +i\frac{\Delta}{2\Omega}\sin\Omega t\right)
\right],\\
c_{11}(t)=e^{-i2\bar\omega t}c_{11}.
\end{cases}
\end{equation}
Here, \(\bar\omega=(\omega_a+\omega_b)/2\) is the mean frequency,
\(\Delta=\omega_b-\omega_a\) is the detuning between the two qubits, and
\(\Omega=\sqrt{(\Delta/2)^2+g^2}\) is the effective Rabi frequency. This unitary evolution over a finite time under the interacting Hamiltonian \(H\) is referred to as a one-step (discrete) collision between the charger and the battery~\cite{CI22}.

\emph{Projective measurement and conditional states}---After the collision of duration \(\tau\), i.e., at \(t=\tau\), the charger is projectively measured in the basis spanned by
\begin{equation}
\label{EQ5}
\begin{cases}
|k_0(\theta)\rangle=\cos\frac{\theta}{2}\,|0\rangle + \sin\frac{\theta}{2}\,|1\rangle,\\
|k_1(\theta)\rangle=-\sin\frac{\theta}{2}\,|0\rangle + \cos\frac{\theta}{2}\,|1\rangle,
\end{cases}
\end{equation}
where \(0\leq\theta\leq\pi\). Then the wave function can be expanded as
\begin{equation}
\begin{aligned}
&|\psi(\tau)\rangle = a_0(\tau) |k_0 0\rangle + a_1(\tau) |k_0 1\rangle + b_0(\tau) |k_1 0\rangle + b_1(\tau) |k_1 1\rangle\\
&= |k_0\rangle\otimes\big(a_0(\tau)|0\rangle + a_1(\tau)|1\rangle\big) + |k_1\rangle\otimes\big(b_0(\tau)|0\rangle + b_1(\tau)|1\rangle\big),
\end{aligned}
\label{EQ6}
\end{equation}
where
\begin{equation}
\label{EQ7}
\begin{cases}
a_0(\tau)=c_{00}(\tau) \cos\frac{\theta}{2} + c_{10}(\tau) \sin\frac{\theta}{2},\\
a_1(\tau)=c_{01}(\tau) \cos\frac{\theta}{2} + c_{11}(\tau) \sin\frac{\theta}{2},\\
b_0(\tau)=-c_{00}(\tau) \sin\frac{\theta}{2} + c_{10}(\tau) \cos\frac{\theta}{2},\\
b_1(\tau)=-c_{01}(\tau) \sin\frac{\theta}{2} + c_{11}(\tau) \cos\frac{\theta}{2}.
\end{cases}
\end{equation}
Thereafter, for simplicity, we drop the explicit \(\tau\)-dependence of \(a_0(\tau),a_1(\tau),b_0(\tau),b_1(\tau)\).
Eq.~\eqref{EQ6} implies that, in general, the charger and the battery become entangled after a collision.

The probabilities of obtaining \(|k_0(\theta)\rangle\) and \(|k_1(\theta)\rangle\) upon measuring the charger are, respectively,
\begin{equation}
p_{k_0} = |a_0|^2 + |a_1|^2, \quad p_{k_1} =1-p_{k_0} = |b_0|^2 + |b_1|^2.
\label{EQ8}
\end{equation}
With probability \(p_{k_0}\), the charger collapses to \(|k_0(\theta)\rangle\), and the battery assumes the corresponding conditional pure state. Without loss of generality, this state can be expressed as
\begin{equation}
|\phi_0\rangle=\frac{a_0|0\rangle + a_1|1\rangle}{\sqrt{p_{k_0}}}
=\sqrt{1-e}|0\rangle+\sqrt{e}\exp(i\alpha)|1\rangle,
\label{EQ9}
\end{equation}
where the excited-state population \(e=|a_1|^2/p_{k_0}\) and \(\alpha=\arg(a_1)-\arg(a_0)\). Note that the global-phase factor is ignored here. With probability \(p_{k_1}\), the charger collapses to \(|k_1(\theta)\rangle\), and the battery assumes
\begin{equation}
|\phi_1\rangle=\frac{b_0|0\rangle + b_1|1\rangle}{\sqrt{p_{k_1}}}
=\sqrt{1-e'}|0\rangle+\sqrt{e'}\exp(i\alpha')|1\rangle,
\label{EQ10}
\end{equation}
where \(e'=|b_1|^2/p_{k_1}\) and \(\alpha'=\arg(b_1)-\arg(b_0)\). Consequently, depending on the measurement outcome, the conditional pure state of the battery \(|\phi_b(\tau)\rangle\) takes one of the two branches: \(|\phi_0\rangle\) or \(|\phi_1\rangle\).

For cascaded collision models~\cite{CI22}, the battery with \(|\phi_b(\tau)\rangle\) further collides with a ``fresh'' charger prepared in $|\phi_a\rangle$ (generally kept fixed) over a time \(\tau\). At the end of this collision step, i.e.,~\(t=2\tau\), a projective measurement is performed on the charger, yielding a new conditional battery state \(|\phi_b(2\tau)\rangle\). This procedure is repeated.

\emph{Adaptive measurements on chargers}---For quantum collision models, the projective measurement basis vectors \(\{|k_0(\theta)\rangle,|k_1(\theta)\rangle\}\) on the charger are often fixed~\cite{CI22}. Generally, for two branches, \(\langle\phi_0|H_\text{b}|\phi_0\rangle\neq\langle\phi_1|H_\text{b}|\phi_1\rangle\), which may be a main contribution to the battery's stored-energy fluctuations over distinct quantum trajectories.

Alternatively, we can choose a suitable projective basis such that the stored energies of the two branches are identical, i.e.,
\begin{equation}
E=\langle\phi_0|H_\text{b}|\phi_0\rangle=\langle\phi_1|H_\text{b}|\phi_1\rangle.
\label{EQ11}
\end{equation}
This is equivalent to
\begin{equation}
\frac{|a_1|^2}{p_{k_0}}=\frac{|b_1|^2}{p_{k_1}}.
\label{EQ12}
\end{equation}
In other words, the excited-state occupation probabilities are the same for the two branches. From Eq.~\eqref{EQ8}, this condition reduces to \(|b_0|^2 |a_1|^2 = |a_0|^2 |b_1|^2\).
For this condition to hold, employing Eq.~\eqref{EQ7}, we derive the constraint on \(\theta\) after the $n$-th collision:
\begin{widetext}
\begin{equation}
\tan{\theta}=\frac{|c_{00}(n\tau)|^2|c_{11}(n\tau)|^2-|c_{10}(n\tau)|^2|c_{01}(n\tau)|^2}
{\mathrm{Re}(c_{01}(n\tau)c_{11}(n\tau)^*)\bigl(|c_{00}(n\tau)|^2+|c_{10}(n\tau)|^2\bigr)
-\mathrm{Re}(c_{00}(n\tau)c_{10}(n\tau)^*)\bigl(|c_{01}(n\tau)|^2+|c_{11}(n\tau)|^2\bigr)}.
\label{EQ13}
\end{equation}
\end{widetext}
From Eq.~\eqref{EQ5}, we can construct the measurement basis \(\{|k_0(\theta)\rangle, |k_1(\theta)\rangle\}\). Since \(\theta\) is adapted to the state \(|\psi(n\tau)\rangle\), we term \(\theta\) the adaptive angle and denote it by \(\theta_a\). This scheme is referred to as an adaptive-angle measurement protocol.

When using this protocol to guide experiments, we can infer the battery conditional states \(|\phi_b(n\tau)\rangle\) from the measurement record \(\{|k(n'\tau)\rangle\}\) with $n'=1,2,\cdots,n$, where each recorded \(|k(n'\tau)\rangle\) is either \(|k_0(\theta)\rangle\) or \(|k_1(\theta)\rangle\). Given a known, fixed pre-collision charger state \(|\phi_a\rangle\), the joint initial state \(|\phi_a\rangle\otimes|\phi_b(n\tau)\rangle\)is fully specified before the next collision. During the collision process, the deterministic dynamical evolution gives the pre-measurement joint state, from which we extract \(c_{00}((n+1)\tau),c_{01}((n+1)\tau),c_{10}((n+1)\tau),c_{11}((n+1)\tau)\). We then solve for \(\theta\) via Eq.~\eqref{EQ13}, construct the measurement basis, and record measurement outcomes for the subsequent collision.

\emph{Stored energies in the battery}---We consider cascaded collision models~\cite {CI22}, in which the chargers are non-interacting and initially uncorrelated, with each charger colliding with the battery exactly once. At the start of every collision, each charger is prepared in the same state
\begin{equation}
|\phi_a\rangle=\sqrt{1-e_a}|0\rangle+\sqrt{e_a}|1\rangle.
\label{EQ14}
\end{equation}
This means the phase of this state is taken as our reference frame. At the beginning of one realization of charging, the battery is prepared in the state \(|\phi_b(0)\rangle=|0\rangle\). Without loss of generality, the conditional pure state of the battery prior to the \({(n+1)}\)-th collision can be written
\begin{equation}
|\phi_b(n\tau)\rangle=\sqrt{1-e_n}|0\rangle+\sqrt{e_n}\exp(i\alpha_n)|1\rangle
\label{EQ15}
\end{equation}
with the stored energy \(E_n=e_n \omega_b\).

For the subsequent collision, the initial joint state reads
\begin{equation}
\psi(n\tau)=|\phi_a\rangle\otimes|\phi_b(n\tau)\rangle.
\label{EQ16}
\end{equation}
After the \((n+1)\)-th collision and subsequent to the measurement with the adaptive angle \(\theta\) on the charger [see Eq.~\eqref{EQ13}], we obtain [see Eq.~\eqref{EQ12}]
\begin{equation}
e_{n+1}=\frac{|a_1|^2}{p_{k_0}}=\frac{|b_1|^2}{p_{k_1}}.
\label{EQ17}
\end{equation}
Here, \(a_1\) and \(b_1\) are the coefficients at \(t=(n+1)\tau\) [see Eq.~\eqref{EQ7}]. Thus we have \(|a_1|^2=e_{n+1} p_{k0}\) and \(|b_1|^2=e_{n+1} p_{k1}\). Since \(p_{k0}+p_{k1}=1\) [see Eq.~\eqref{EQ8}], we obtain \(e_{n+1}=|a_1|^2+|b_1|^2\). This is an algebraic identity rather than a physical superposition of the two post-selected branches. According to Eq.~\eqref{EQ7}, \(|a_1|^2+|b_1|^2
=|c_{01}((n+1)\tau)|^2+|c_{11}((n+1)\tau)|^2\). The coefficients \(c_{01}((n+1)\tau)\) and \(c_{11}((n+1)\tau)\) can be calculated using Eq.~\eqref{EQ4}, with the initial joint state for the \((n+1)\)-th collision given in Eq.~\eqref{EQ16} as input. Therefore,
\begin{equation}
\begin{aligned}
&e_{n+1}=\\
&e_a e_n+\left(C_\tau^2+\frac{\Delta^2}{4\Omega^2}S_\tau^2\right)(1-e_a)e_n
+\frac{g^2}{\Omega^2}S_\tau^2 e_a(1-e_n)\\
&+2\frac{g}{\Omega}S_\tau\sqrt{e_a(1-e_a)e_n(1-e_n)}
\left(\frac{\Delta}{2\Omega}S_\tau\cos\alpha_n-C_\tau\sin\alpha_n\right),
\end{aligned}
\label{EQ18}
\end{equation}
where \(C_{\tau}=\cos(\Omega\tau)\) and \(S_{\tau}=\sin(\Omega\tau)\).

\emph{Results and discussions}---In simulations, \(10^6\) stochastic quantum trajectories, i.e., charging realizations, are sampled. Additional trajectories yield the same averaged quantities.

\emph{Complete suppression of stored-energy fluctuations}---Eq.~\eqref{EQ14} gives the general state of the charger qubit, namely \(|\phi_a\rangle=\sqrt{1-e_a}|0\rangle+\sqrt{e_a}|1\rangle\). When each fresh charger qubit is prepared in the excited state, i.e., \(e_a=1\), Eq.~\eqref{EQ18} reduces to
\begin{equation}
e_{n+1}=e_n+\frac{g^2}{\Omega^2}S_\tau^2 (1-e_n).
\label{EQ19}
\end{equation}
At $t=0$, the battery state is \(|\phi_b(0)\rangle=|0\rangle\), so $e_0=0$. The corresponding stored energy reads
\begin{equation}
 E_{n+1}=e_{n+1} \omega_b=[(1-\frac{g^2}{\Omega^2}S_\tau^2)e_n+\frac{g^2}{\Omega^2}S_\tau^2]\omega_b.
\label{EQ20}
\end{equation}
Note that \(e_{n+1}\) is independent of the phase \(\alpha_n\) in \(|\phi_b(n\tau)\rangle\) [see Eq.~\eqref{EQ15}].
With the adaptive angle \(\theta_a\) [see Eq.~\eqref{EQ13}], we can obtain the two conditional pure states \(|\phi_0\rangle\) and \(|\phi_1\rangle\) of the battery [see Eqs.~\eqref{EQ5}--\eqref{EQ10}]. They take the form
\begin{equation}
\label{EQ21}
\begin{cases}
|\phi_0\rangle=\sqrt{1-e_{n+1}}|0\rangle+\sqrt{e_{n+1}}\exp(i\alpha_{n+1}^0)|1\rangle,\\
|\phi_1\rangle=\sqrt{1-e_{n+1}}|0\rangle+\sqrt{e_{n+1}}\exp(i\alpha_{n+1}^1)|1\rangle.
\end{cases}
\end{equation}
In the next iteration, each branch (\(|\phi_b((n+1)\tau)\rangle=|\phi_0\rangle\) or \(|\phi_1\rangle\)) undergoes collision and adaptive measurement. Eq.~\eqref{EQ19} shows that \(e_{n+1}\) is independent of the phase \(\alpha_n\). For the same reason, \(e_{n+2}\) is also independent of the phases \(\alpha_{n+1}^0\) and \(\alpha_{n+1}^1\). This procedure yields four possible descendant branches with identical stored energies, given by
\begin{equation}
E_{n+2}=e_{n+2} \omega_b=\left[(1-\frac{g^2}{\Omega^2}S_\tau^2)e_{n+1}+\frac{g^2}{\Omega^2}S_\tau^2\right]\omega_b.
\label{EQ22}
\end{equation}
By mathematical induction, all trajectory branches after the \(\tilde{n}\)-th collision possess strictly equal battery energies. This implies that stored-energy fluctuations are completely suppressed, i.e., identical stored energies are achieved.

\begin{figure}[!htbp]
\centering
\includegraphics[width=0.48\columnwidth]{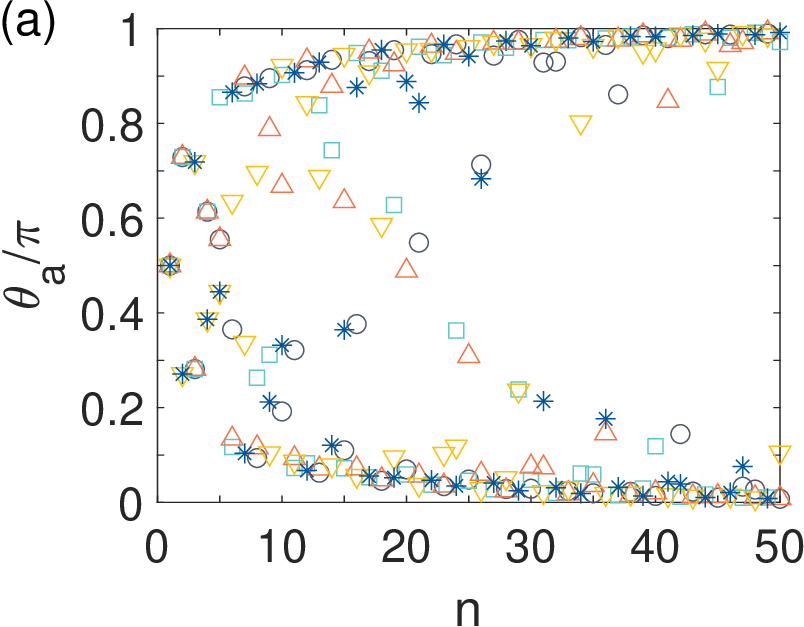}
\hspace{4pt}
\includegraphics[width=0.405\columnwidth]{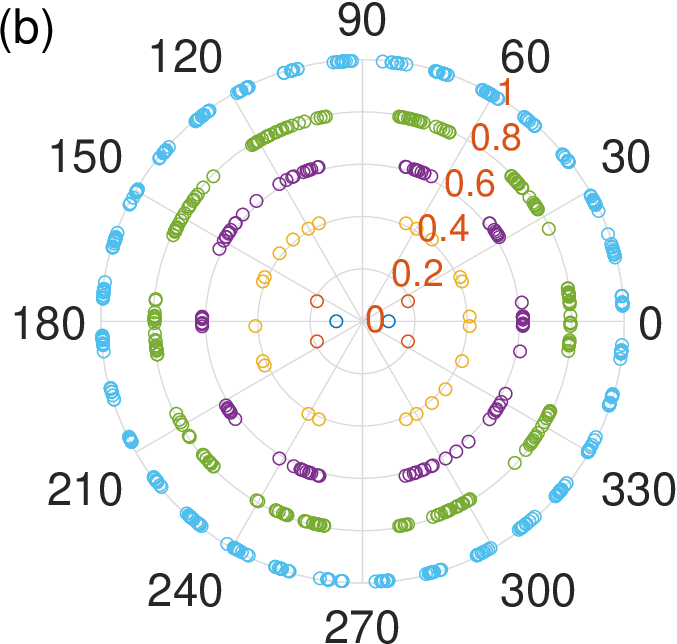}
\vspace{0.2cm}

\includegraphics[width=0.48\columnwidth]{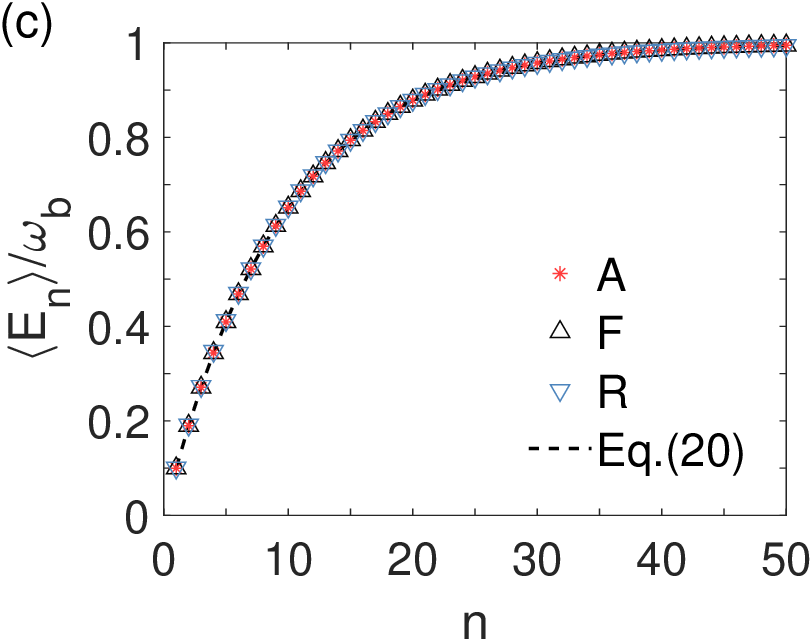}
\hspace{4pt}
\includegraphics[width=0.48\columnwidth]{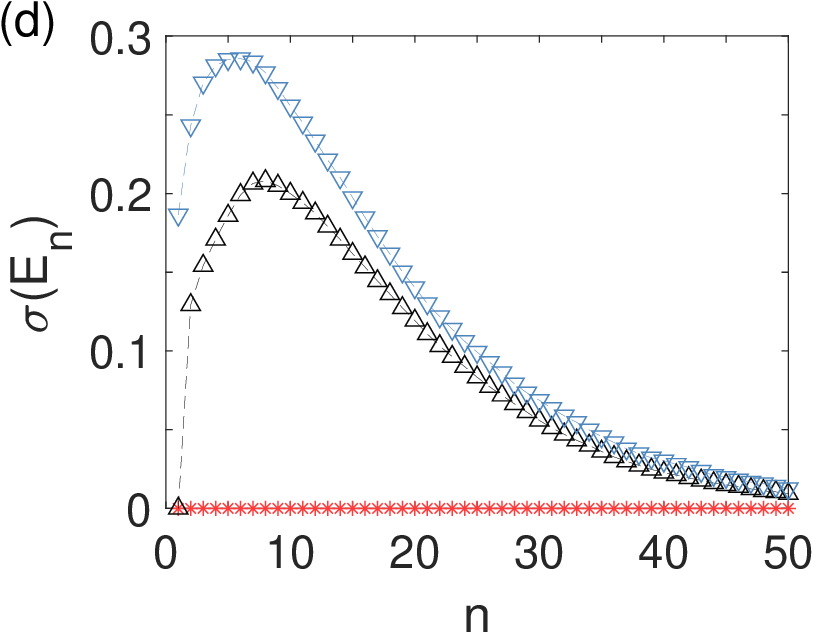}
\caption{Results for charger qubits in the excited state \(|\phi_a\rangle=|1\rangle\). (a) The adaptive angle \(\theta_a\) versus collision steps \(n\) (five quantum trajectories denoted by five different symbols). (b) Conditional pure state of the battery \(|\phi_b(\tau)\rangle=\sqrt{1-e_n}|0\rangle+\sqrt{e_n}\exp(i\alpha_n)|1\rangle\) for \(n=1,2,5,9,15,50\) (\(1000\) quantum trajectories); polar angle \(\alpha_n\), and radius \(e_n\) increases outward. (c) Mean stored energies \(\langle{E_n}\rangle\) and (d) stored-energy fluctuations \(\sigma(E_n)\) versus \(n\) for measurement protocols with adaptive (A), fixed (F) and random (R) angles. Parameters: \(\Delta=0.3\), \(g=0.05\) and \(\Omega\tau=\pi/2\); \(\theta_f=\pi/2\).}\label{Fig1}
\end{figure}

At the detuning \(\Delta=\omega_b-\omega_a=0\), the effective Rabi frequency \(\Omega=\sqrt{(\Delta/2)^2+g^2}=g\). For \(\Omega\tau=\pi/2\), Eq.~\eqref{EQ19} gives \(e_{1}=1\), meaning full charging of the battery in a single collision step. This case recovers the standard Rabi model. In this Letter, we focus on cases away from resonance.

For a visual demonstration of the cascaded-collision process, Fig.~\ref{Fig1}(a) illustrates the evolution of adaptive angles \(\theta_a\) along successive collision steps. It shows that \(\theta_a\) takes different values in distinct quantum trajectories, which reflects the randomness originating from quantum measurements. Fig.~\ref{Fig1}(b) shows that after the n-th collision, \(e_{n}\) are the same for distinct quantum trajectories, although angles \(\alpha_n\) are different. This has been analyzed right after Eq.~\eqref{EQ21}, where we reveal that variations in angles \(\alpha_n\) do not affect the value of the occupation probability \(e_{n+1}\). Consequently, all trajectory branches after the \(n\)-th collision possess strictly equal values of \(e_{n+1}\) for the battery.

Fig.~\ref{Fig1}(c) demonstrates that the curves for the mean stored energy \(\langle E_n\rangle\) at each collision step, averaged over all trajectories, are identical under the adaptive-, fixed-, and random-angle measurement protocols. As discussed in Ref.~\cite{CI22}, averaged quantities over all stochastic trajectories are independent of the choice of ancilla measurement basis \(\{|k_0(\theta)\rangle, |k_1(\theta)\rangle\}\). Furthermore, these \(\langle E_n\rangle\) values agree with the predictions of Eq.~\eqref{EQ20}. In contrast, Fig.~\ref{Fig1}(d) shows that the corresponding stored-energy fluctuations \(\sigma(E_n)\) differ among the three protocols. In particular, fluctuations are completely suppressed in our adaptive-angle measurement protocol. Here, the fixed-angle \(\theta_f\) is optimized to minimize the mean fluctuations \(\langle\sigma(E_n)\rangle_f\), averaged over all collision steps, whereas the random-angle \(\theta_r\) is drawn independently and uniformly from \([0,\pi]\) at each collision step.

\begin{figure}[!htbp]
\centering
 \includegraphics[width=0.48\columnwidth]{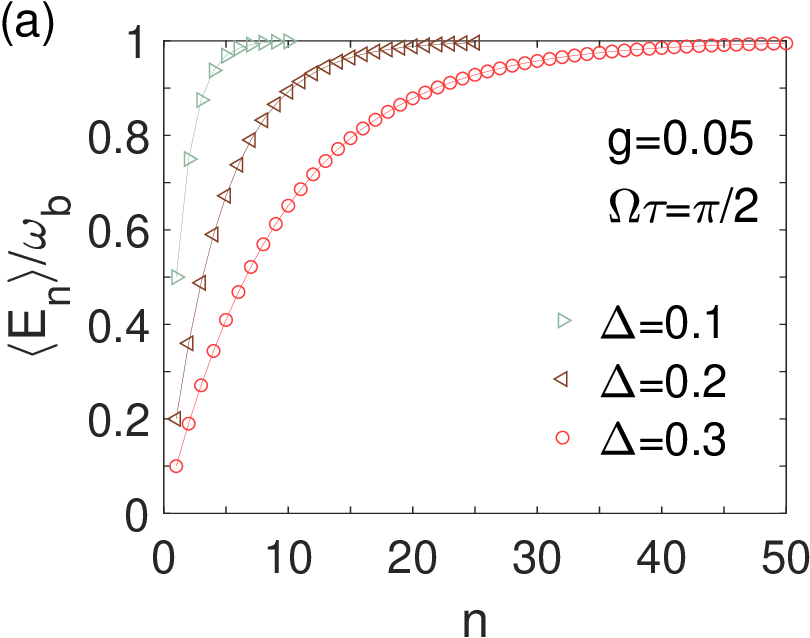}
\vspace{0.2cm}

 \includegraphics[width=0.48\columnwidth]{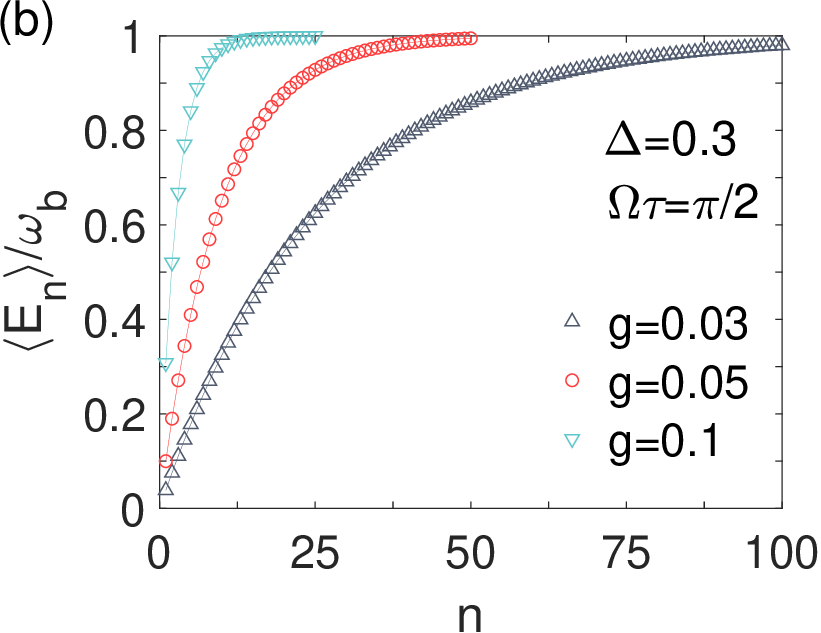}
 \hspace{4pt}
 \includegraphics[width=0.48\columnwidth]{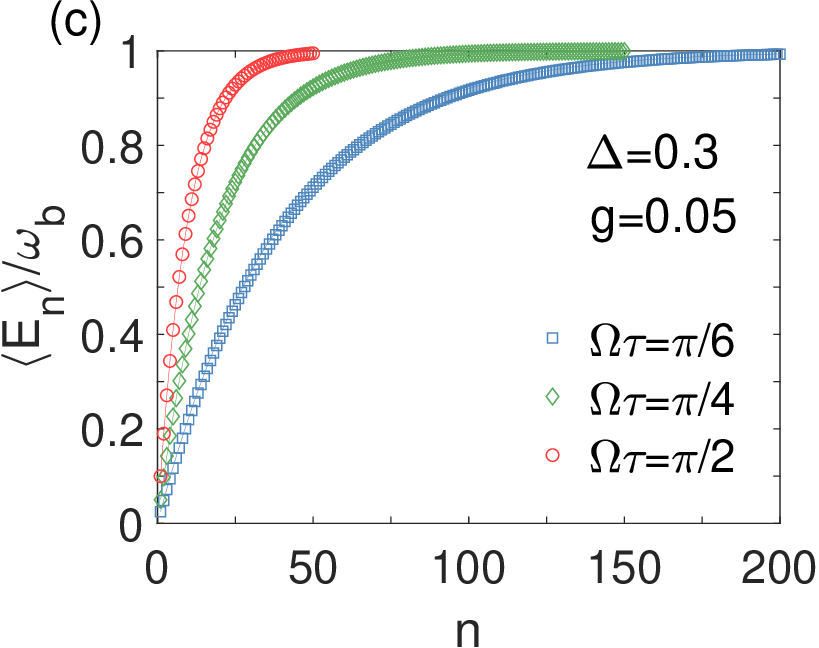}
\caption{For the adaptive-angle measurement protocol, mean stored energies \(\langle E_n \rangle\) versus collision steps \(n\) for charger qubits prepared in the excited state \(|\phi_a\rangle=|1\rangle\). Panels (a)--(c) correspond to different parameter sets. Lines denote theoretical predictions from Eq.~\eqref{EQ20}.} \label{Fig2}
\end{figure}

Fig.~\ref{Fig2} shows mean stored energies \(\langle{E_n}\rangle\) versus collision steps \(n\) for different parameter sets. All numerical simulation results agree with theoretical predictions given in Eq.~\eqref{EQ20}. In all cases, stored-energy fluctuations \(\sigma(E_n)\) are zero and hence not displayed.  Thus, our proposed protocol remains effective for complete fluctuation suppression over a wide range of parameters.

\begin{figure}[!htbp]
\centering
 \includegraphics[width=0.405\columnwidth]{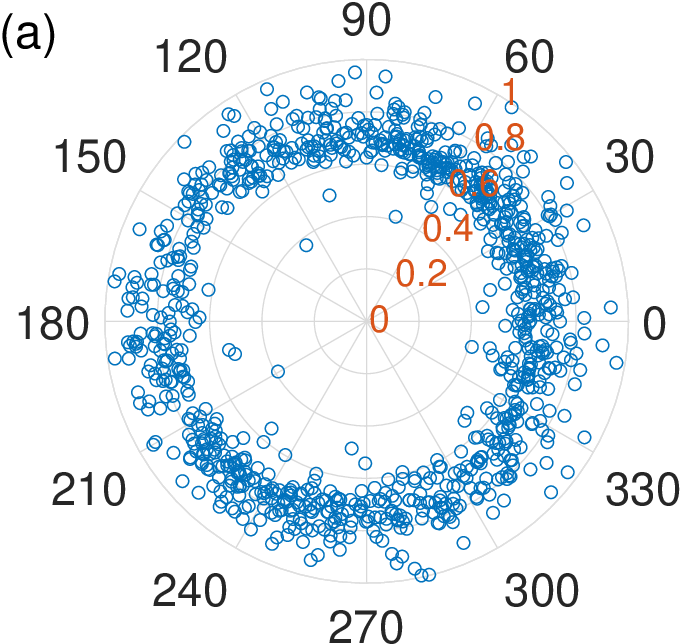}
 \hspace{4pt}
 \includegraphics[width=0.48\columnwidth]{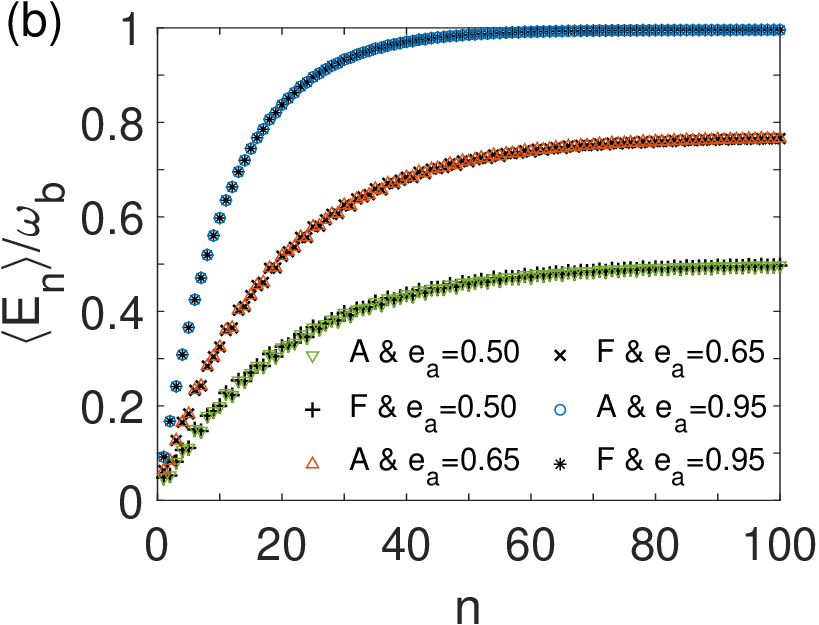}
 \vspace{0.2cm}

 \includegraphics[width=0.48\columnwidth]{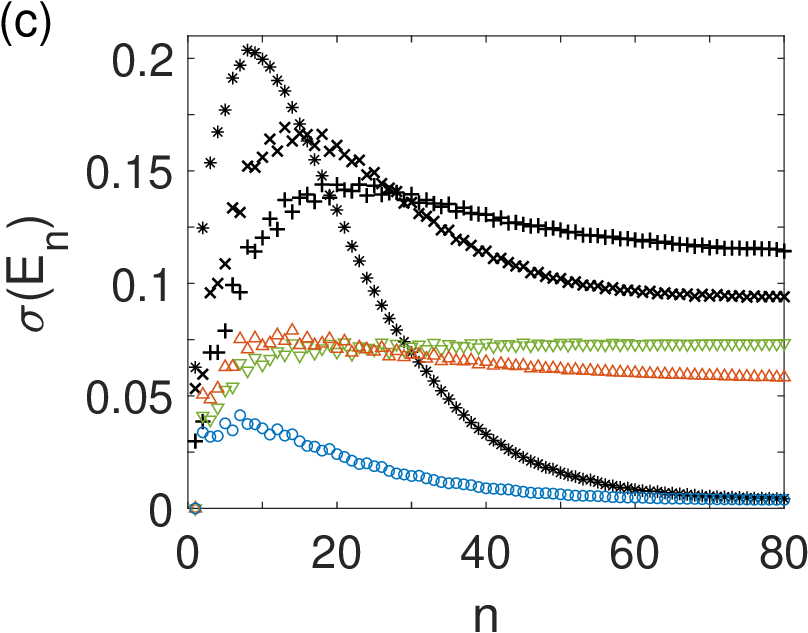}
 \hspace{4pt}
 \includegraphics[width=0.48\columnwidth]{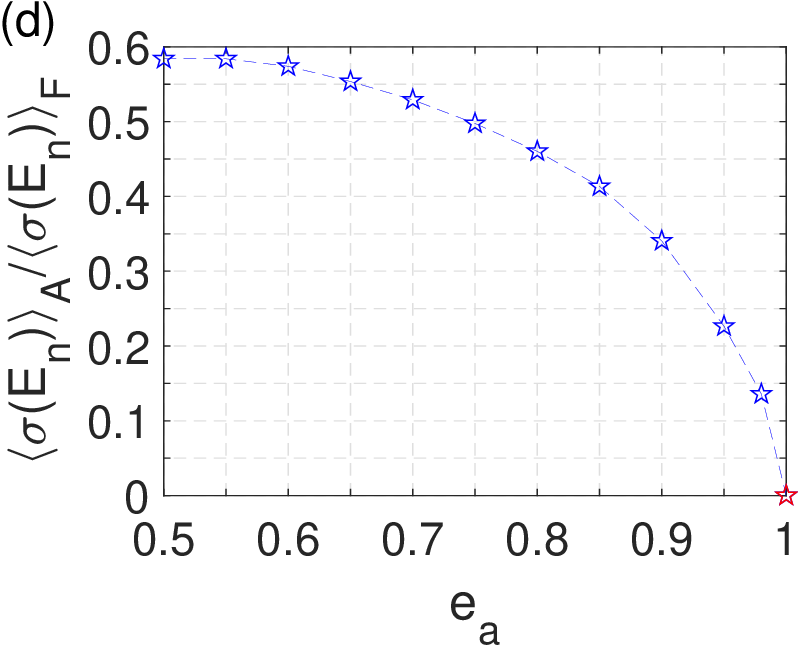}
\caption{Results for charger qubits in the superposition state \(|\phi_a\rangle=\sqrt{1-e_a}|0\rangle+\sqrt{e_a}|1\rangle\). (a) Conditional pure state of the battery \(|\phi_b(\tau)\rangle=\sqrt{1-e_n}|0\rangle+\sqrt{e_n}\exp(i\alpha_n)|1\rangle\) from 1000 quantum trajectories for \(e_n=0.65\) and \(n=60\). (b) Mean stored energies \(\langle E_n \rangle\) and (c) stored-energy fluctuations \(\sigma(E_n)\) versus \(n\), where A and F denote adaptive-angle and fixed-angle measurement protocols, respectively. (d) Ratio of average stored-energy fluctuations versus \(e_a\). Parameters: \(\Delta=0.3\), \(g=0.05\), \(\Omega\tau=4\pi/9\); \(\theta_f=0.1\) for \(e_a=0.5, 0.55, \cdots,0.70\), and \(\theta_f=0.15, 0.2, 0.2, 0.25, 0.3, 0.4\) for \(e_a=0.75, 0.80, \cdots,0.95, 0.98\).} \label{Fig3}
\end{figure}

\emph{Substantial suppression of stored-energy fluctuations}---The charger qubit is generally prepared in the state \(|\phi_a\rangle=\sqrt{1-e_a}|0\rangle+\sqrt{e_a}|1\rangle\), as given by Eq.~\eqref{EQ14}. If \(e_a\neq0\) and \(1\), it represents a superposition state. In this case, Eq.~\eqref{EQ18} shows that \(e_{n+1}\) depends on the phase \(\alpha_n\) in \(|\phi_b(n\tau)\rangle\) [see Eq.~\eqref{EQ15}]. Since the two conditional states \(|\phi_0\rangle\) and \(|\phi_1\rangle\) may carry different phases \(\alpha_{n+1}^0\) and \(\alpha_{n+1}^1\) [see Eq.~\eqref{EQ21}], the subsequent four descendant branches will generally yield different values for \(e_{n+2}\). By mathematical induction, distinct battery energies generally develop among trajectory branches after the \(n\)-th collision. This implies that stored-energy fluctuations exist. Fig.~\ref{Fig3}(a) illustrates this theoretical consequence, where dispersion in the values of \(e_{n}\) is observed.

For comparison, Fig.~\ref{Fig3}(b) shows the mean stored energies \(\langle E_n\rangle\) under the adaptive-angle and fixed-angle protocols, obtained by averaging over all stochastic trajectories. They are independent of the choice of protocol. In contrast, Fig.~\ref{Fig3}(c) shows that the stored-energy fluctuations \(\sigma(E_n)\) do depend on it. To quantify this difference, we define the ratio \(\eta=\langle\sigma(E_n)\rangle_A/\langle\sigma(E_n)\rangle_F\). Figure~\ref{Fig3}(d) shows that \(\eta\) decreases continuously from $0.5840$ to $0$ as \(e_a\) increases from $0.5$ to $1$. The endpoint \(e_a=1\) corresponds to the charger qubit in the excited state \(|\phi_a\rangle=|1\rangle\), i.e., the case of \emph{complete suppression of stored-energy fluctuations} discussed above. Therefore, even for a superposition-state charger qubit, our proposed adaptive-angle protocol is superior to the fixed-angle protocol in suppressing stored-energy fluctuations.

\emph{Conclusions}---This work establishes an adaptive-angle measurement paradigm for collisional quantum battery charging, in which post-collision measurement bases for chargers are dynamically tuned to suppress trajectory-dependent energy disparities. For each ``fresh'' charger prepared in the excited state, stored-energy fluctuations across different quantum trajectories are fully eliminated. For superposition-state chargers, substantial reduction of stored-energy fluctuations is achieved relative to the minimal-fluctuation fixed-measurement basis. The proposed protocol maintains its performance across a broad range of detunings. This adaptive scheme provides an effective strategy for achieving high-uniformity quantum battery charging. It may inspire experimental realizations as well as further developments of quantum-battery designs.

\emph{Acknowledgments}---This work was supported by the National Natural Science
Foundation of China (Grant No. 12674301).

\emph{Data availability}---There are no publicly available
research data or software supporting this manuscript.
Requests for further information or data should be sent
to the authors.

\end{document}